# Implementation of Web-Based Respondent-Driven Sampling among Men who Have Sex with Men in Vietnam

*Linus Bengtsson[1*], Xin Lu[1,2], Quoc Cuong Nguyen[3], Martin Camitz[1], Nguyen Le Hoang[4], Fredrik Liljeros[2], Anna Thorson[1]*

1) Karolinska Institutet, Dept. of Public Health Sciences, Stockholm, Sweden; 2) Stockholm University, Dept. of Sociology, Stockholm, Sweden; 3) Family Heath International, Hanoi, Vietnam; 4) iSEE, Hanoi, Vietnam

**Abstract**

**Objective:** *Lack of representative data about hidden groups, like men who have sex with men (MSM), hinders an evidence-based response to the HIV epidemics. Respondent-driven sampling (RDS) was developed to overcome sampling challenges in studies of populations like MSM for which sampling frames are absent. Internet-based RDS (webRDS) can potentially circumvent limitations of the original RDS method. We aimed to implement and evaluate webRDS among a hidden population.*

**Methods and Design**: *This cross-sectional study took place 18 February to 12 April, 2011 among MSM in Vietnam. Inclusion criteria were men, aged 18 and above, who had ever had sex with another man and were living in Vietnam. Participants were invited by an MSM friend, logged in, and answered a survey. Participants could recruit up to four MSM friends. We evaluated the system by its success in generating sustained recruitment and the degree to which the sample compositions stabilized with increasing sample size.*

**Results:** *Twenty starting participants generated 676 participants over 24 recruitment waves. Analyses did not show evidence of bias due to ineligible participation. Estimated mean age was 22 year and 82% came from the two large metropolitan areas. 32 out of 63 provinces were represented. The median number of sexual partners during the last six months was two. The sample composition stabilized well for 16 out of 17 variables.*

**Conclusion:** *Results indicate that webRDS could be implemented at a low cost among Internet-using MSM in Vietnam. WebRDS may be a promising method for sampling of Internet-using MSM and other hidden groups.*

**Key words:** *Respondent-driven sampling, Online sampling, Men who have sex with men, Vietnam, Sexual risk behavior*

---

[*]*Corresponding author: Linus Bengtsson, Division of Global Health/IHCAR, Department of Public Health Sciences, Karolinska Institutet, Nobels väg 9, SE-171 77, Stockholm, Sweden, Mobile: +46 (0) 707 507578, bengtssonlinus@gmail.com, linus.bengtsson@ki.se*

# Introduction

Men who have sex with men (MSM) has emerged as a key population in the global HIV epidemic [1-3]. Modeling work on the Asian epidemic points to a scenario in which 42 percent of all new HIV infections in Asia will occur among MSM by 2020 [4]. While population-based surveys in countries with generalized epidemics have generated vast amounts of data on sexual behavior [5], studies on MSM and other hidden populations struggle to generate representative samples and adequate sample sizes [1]. The lack of representative data of MSM risk-behavior severely hinders an understanding of the underlying dynamics of the MSM epidemics and prevents an evidence-based response. New cost-efficient methods for representative sampling of MSM and other hidden groups are thus needed.

Respondent-driven sampling (RDS) was developed to overcome sampling challenges in studies of populations for which a sampling frame is difficult or impossible to define, such as MSM, injecting drug users (IDU), and sex workers (SW) [6-11]. An RDS study starts by purposively selecting a handful of participants who are known members of the study population. These "seeds" are given invitation coupons (usually three) to distribute to other members of the population. These members are in turn given three new coupons to distribute. Monetary incentives are usually given both for participation and recruitment.

RDS resembles "snowball sampling" [12] but differs from it in several important respects. The incentive system of RDS allows the creation of long recruitment chains. If the sampling conforms to methodological assumptions, the proportion of the sample with a certain characteristic stabilizes at a level determined by the characteristics of the population, independently of the characteristics of the seeds [7]. Furthermore, snowball sampling systematically oversamples individuals with many contacts. All individual properties correlated with the number of contacts within the group under study will hence be over or under sampled in a snowball sample. In contrast, during an RDS study researchers record an estimate of each person's social network size and adjust for this bias. Participants are also not required to name or identify their contacts, as is often the case in snowball sampling. Instead participants can pass invitation coupons to any of their contacts at their own discretion. They receive a reward when their contacts participate in the survey, serving to increase participation rates and decrease selection bias. The RDS method has been shown to be analytically unbiased under a limited number of assumptions. It is currently the focus of extensive methodological research [9, 13-17].

Although RDS in certain contexts has clear advantages over other sampling methods [18], the standard implementation of the method has several limitations, including: 1) individuals with a behavior that is stigmatized, illegal, or associated with high privacy concerns, may be unwilling to access survey offices physically and may thus be underrepresented in the sample; 2) persons from middle- and upper income levels may not be sufficiently incentivized by the study rewards, given the time and effort required to participate ; 3) the geographic area of study needs to be small enough to allow participants to travel to the study sites; and 4) RDS studies are, like other field survey methods, relatively expensive since they require the presence of trained staff for extended periods of time and need to be repeated at several sites to generate national or regional data,

Sampling participants though the Internet can mitigate some of these disadvantages by allowing people to participate anonymously and with little effort. Online sampling also allows for vast geographic coverage and may be carried out at markedly lower cost than standard field surveys. However, current methods of Internet-based sampling of hidden groups enroll participants through self-selection, which may cause important bias [19]. Usually, a banner add is put on a web page, e.g. for gay men and persons accessing this site then click the banner to volunteer for the study. These surveys can have participation rates as low as a few in a thousand to a few in a hundred out of registered users [20, 21].

Web-based RDS (henceforth webRDS) can potentially circumvent both the disadvantages of standard RDS as well as those inherent in current Internet-based sampling methods of hidden groups. To our knowledge, there are only two published webRDS surveys, both involving students at Cornell University. The results of these studies showed that the RDS estimates agreed relatively well with the true characteristics of the student population [16]. These surveys did not, however, target a hidden population, and they did not include questions with sensitive content. We aimed to implement and evaluate webRDS for sampling and surveying of a hidden and stigmatized population, Internet-using MSM in Vietnam.

## Materials and Methods

*General Study Design:* The survey was cross-sectional, performed online and carried out between February 18 and April 12, 2011, applying web-based respondent driven sampling (webRDS).

*Inclusion criteria:* Eligible participants were adult men (18 years and above) who had ever had sex with another man, had not previously participated in the survey, and were living in Vietnam at the time of the study.

*MSM and Internet use in Vietnam:* Internet access in Vietnam costs approximately 0.15 USD per hour at Internet cafés. MSM in Vietnam are stigmatized [22], and HIV prevalence in the group has been estimated at 14-20% and 14-16% in Hanoi and Ho Chi Minh City, respectively (2009) [23]. Internet use as a proportion of the population in Vietnam was 27% in 2010 (24 million persons) and 60% and 50%, respectively, in the large urban areas of Hanoi and Ho Chi Minh City [24]. Internet use among MSM in general may be considerably higher than in the general population [25]. Ninety-four percent of MSM in an offline RDS in Hanoi stated that they used the Internet [26].

*Sampling:* The study was performed in collaboration with a local research organization in Vietnam (iSEE), which has an extensive contact network among MSM community groups. Fifteen seeds, who were recruited through these networks, initiated the survey and a further five seeds were added two weeks later to increase the speed of recruitment. Six seeds came from Ho Chi Minh City, ten from Hanoi and four from Hoà Bình. Nineteen out of the 20 seeds had attended some kind of education after high school (vocational training, college or university). Participants received an invitation message with a login code and a web address from their recruiter. They logged in, approved participation and eligibility, and answered a questionnaire. Participants who wanted to recruit MSM friends provided an e-mail or Yahoo! Messenger address (popular for communications in Vietnam) and were automatically sent four invitation messages to forward to MSM friends. Reminders to recruit were sent out two and four days after completing the survey.

*Web site:* The graphic design of the web site aimed at giving a friendly impression without strong MSM connotations.

*Incentives and recruitment stimuli:* 1) text emphasizing participation in order to support MSM in Vietnam; 2) 2.45 USD (50,000 VND) as credit on the participant's SIM card and the same amount for each successful recruitment of an MSM friend (maximum four); 3) the option of donating the monetary reward to an MSM community organization chosen by the participant; 4) a lottery with the possibility of winning an iPad; and 5) being able to compare one's own answers to those of other participants in simple, informative and anonymous charts. Eight survey questions were included specifically to stimulate the participants' interest in comparing themselves with other participants.

*Earlier versions of the system:* The web site and recruitment system was extensively pilot tested. Two versions of the webRDS site was used for sampling before the study described in this paper was carried out. These webRDS systems differed in that they had a less advanced graphic design and smaller incentives. In the first survey, recruitment died out after a maximum of 5 waves (25 participants, 15 seeds). The second time, recruitment improved but stopped after 5 waves (84 participants, 15 seeds).

*Data collection:* The questionnaire contained 17 questions covering sexual partner numbers during the last 6 months, sexual partner preferences, the duration of the respondent's longest relationship, opinions on same-sex marriage in Vietnam, frequency of Internet use, socio-demographic characteristics, network size, and relationship between the participant and his recruiter. Participants who wanted to receive rewards filled out contact details and a personal identifier (telephone number, email or Yahoo! Messenger address, and the last three digits of their nine-digit ID number). Time points at which each participant loaded the web pages was stored to facilitate identification of ineligible submissions, including unserious attempts to answer the questionnaire or the same person trying to answer more than one questionnaire to receive additional rewards.

*Analyses of duplicated submissions, data cleaning and analysis:* 9.6% of completed surveys (65 surveys) included a stated age below 18 years, or a telephone number, e-mail or Yahoo! Chat address that had previously been registered in the system. We defined these as "invalid". We excluded seeds (customary in RDS analysis [8]) together with the aforementioned invalid submissions to produce a cleaned sample. From this sample we estimated, in Matlab, population proportions using the current state of the art estimator, RDSII, which requires only information on the sample compositions and the social network sizes of the participants [10].
We checked all surveys for other signs of duplication or invalidity by flagging surveys containing a repeated IP number, deviating answers, or short completion times. We analyzed the sensitivity of the estimates to inclusion and exclusion of these flagged submissions. Specifically we compared the RDS II estimates generated from the full sample of non-seed submissions with valid age with the RDS II estimates generated from groups with progressively stricter inclusion criteria according to the following: 1) exclusion of submissions with a repeated email, Yahoo! Chat ID or telephone number (forming the cleaned sample above); 2) additionally excluding repeated IP numbers; and 3) additionally excluding submissions with short completion times (< three minutes), submissions stating no education (rare in Vietnam), or submission stating six-month partner numbers above 1,000. The average absolute differences in estimates when comparing the full sample to these groups with progressively stricter inclusion criteria were less than 0.64% (maximum difference 6.6%). Additional details of these group comparisons are included in the supplementary material.

*Personal network size:* We asked participants for the number of MSM they had interacted with in any way during the past seven days (including on the phone, Internet, or in person) and how many of these persons they believed used the Internet. We used these indicators to define the participants' personal network sizes. We replaced missing personal network size data with the RDSII-estimated average network size from submissions with non-missing network data. The average network size was 5.5 persons. We compared these RDSII estimates with RDS II estimates where we excluded submissions with missing network size data. The differences were small (see supplementary material for detailed comparisons).

*Evaluation and analyzes of equilibrium:* As there is no gold standard by which to validate the sampling, we evaluated the system in terms of its success in generating sustained recruitment, the degree to which the sample compositions stabilized with increasing sample size (independence of the sample from the seeds), and finally, in the discussion, we contrast the sample compositions with results from other surveys.
We analyzed whether equilibrium was achieved in two ways. We first used the standard criterion from the RDS literature [27]. This criterion requires the sampling process to have reached a certain number of waves. The number of waves required is, for each variable, determined by the number of steps required by a first-order Markov process to reach a less than a two percent relative difference between its value at a given step and its value after an infinite number of steps. The transition probabilities used to calculate the values of the Markov process are the averaged transition probabilities in the study's recruitment chains. Second, we produced plots of the changes in the sample compositions as sample size increased.

*Ethics:* IP addresses were converted to a unique anonymous code using a one-way encryption algorithm, and the original IP numbers were deleted. Login passwords were only valid for a single session, and communication between the users and the server was encrypted.

# Results

**Recruitment Dynamics**

676 study participants submitted a survey during the study period. The length of recruitment trees varied from 1 to 24 waves (excluding seed wave). Eight recruitment trees (out of 20) reached more than five waves (Fig. 1a)

Fig. 1a

Five seeds were added 14 days after the first group (see methods). For clarity of presentation we backdated the start date of these five seeds 14 days so that all seeds could be considered to have started on the same day. Using this adjustment, the site received slightly less than 500 submissions during the first two weeks of sampling. The daily number of submissions then gradually decreased and about 100 surveys were submitted during the last 20 days, after which submissions stopped (Fig. 1b).

Fig. 1b

**Equilibrium**

Using the standard criteria in the literature [27], equilibrium was reached for all variables after a maximum of seven waves and a median of two waves. We also plotted the sample compositions with increasing sample sizes. Selected variables are shown in Fig. 2 and all variables are available in the supplementary material. Judging from these plots, the sample compositions stabilized well for all variables in the survey, with the exception of home province. The mean change in samples compositions comparing the full sample and the sample excluding the last 200 participants was for all other estimates 1.5% for proportional estimates and 0.32 for numeric estimates (sexual partners, age, MSM friends, MSM friends with Internet). The maximum changes for proportional estimates and numerical estimates were 2.3% and 0.035 respectively.

Fig. 2 (a-d)

**Characteristics of the sample**

The majority of the sample consisted of young persons with an estimated mean and median age of 22 years. A high proportion had some education after high school (estimated proportion with education at vocational school or college: 87%). An estimated 67% used the Internet every day during the past month and an estimated 82% came from the two large metropolitan areas of Ho Chi Minh City and Hanoi (81% of the sample). The recruitment trees also penetrated outside the large metropolitan areas with 32 provinces represented out of 63 (Fig. 3a-c).

Fig. 3 (a-c)

An estimated 98% (99% of the sample) preferred only men or preferred men to women as sexual partners, and 81% (81% of the sample) thought that same-sex marriage should be allowed in Vietnam. An estimated 92% (91% of the sample) had a pre-existing social relationship with their recruiter. Median number of sexual partners during the last six months was two (Fig. 4a-c).

Fig. 4 (a-c)

**Discussion**

We developed an automatic webRDS system to sample men who have sex with men (MSM) in Vietnam, a country in which same sex relationships are highly stigmatized and can lead to severe consequences if revealed to family members or colleagues (19). We successfully used the system to sample and survey 676 MSM on a number of sensitive issues. We evaluated the independence of the seeds from the sample by showing that sample composition stabilized very well for all variables, possibly with the exception of home province. We used a varied set of incentives to stimulate participation and recruitment, which became rapid and robust, reaching 24 waves. The results indicate that webRDS could potentially be implemented at a low cost among Internet-using MSM as well as potentially becoming a valuable method for sampling other Internet-using populations.

Comparing available national statistics with the study estimates yields interesting similarities and dissimilarities that may reflect differences between the sexually active Internet-using MSM population and the general population or may reflect bias in sampling. Income distributions (Fig. 3b) is broadly consistent with the national average per capita income for urban areas (2,130,000 VND, 2010 [28]). Using the RDSII estimator, 97% of the MSM population under study was estimated to be below 30 years of age and the sample mean and median ages were 22 years. By comparison, 43% of the adult male population in Vietnam is between 18 and 29 [29]. The lower mean age of sampled MSM compared to the national age distributions for men is broadly consistent with other offline MSM samplings in Vietnam, which reported a median age of 24 years [30]) and a mean age of 27 years [20, 31]. An estimated 99% of MSM in our sample had completed upper secondary school as compared to 93% nationally [32]. The sample was heavily concentrated to the two large metropolitan areas of Ho Chi Minh City and Hanoi, with a population estimate of 84% for these cities combined. These cities constitute approximately 55% of the urban population (2009 and 2010) [33, 34] in Vietnam and about 16% of the whole population [33]. Primarily migration of young MSM as well as possibly different levels of access to the Internet may form part of the explanation for the observed differences. The recruitment chains in our sample did frequently crossed over between provinces. In total, 29.8% of all recruitment events took place between persons in different provinces. We did thus not find evidence for that the men's social networks formed geographically isolated groups, which otherwise would have been a source of bias. Additionally, like other social networks, MSM networks in Vietnam are most likely small-world networks [35], with short numbers of steps between provinces.

There are a few limitations to this study. We excluded 13.0% (n=85) of the submissions because of duplicated personal information or an age below 18 years. While this shows that the recruitment system did not work perfectly, it also shows the potential for eliminating duplicate submissions. 17.5% of completed surveys (115 surveys) included an IP number that had previously been registered in the system, which may signal duplicated submissions. However, it is important to note that IP-numbers are shared by all users at an Internet cafe and often by all users within a neighborhood. An array of non-Vietnamese IP-numbers is also used in Vietnam to access restricted sites like Facebook. We checked whether the final estimates were sensitive to exclusion of these submissions as well as of submission with very fast completion time, and did not find that this was the case. Similar protocols for quality check as those used in this study have been employed in other Internet-based surveys among MSM [36]. Because sincere and insincere participants are likely to interact differently with web survey pages, analyzes of online behavioral data gathered during surveys may in the future provide a way to further improve these protocols.

Although this study aimed to sample Internet-using MSM, access to the Internet will be a limiting factor for representative sampling of MSM in general, using webRDS. Older, poorer, and rural persons may be undersampled. However, Internet penetration among MSM in Vietnam may be very high, specifically among urban MSM and in particular in metropolitan areas, and 94% of MSM in an offline RDS study in Hanoi stated that they used the Internet [26].

Although more than 600 submissions were received within 5 weeks, recruitment eventually died out despite being far from the total size of the Internet-using MSM population in Vietnam, which is probably above 10,000. The most likely explanation for why the survey died out is not global saturation but local saturation. Fig. 1b shows the typical functional form of a diffusion process reaching its carrying capacity [37]. It is very common in acquaintance networks that an individual's neighbors are connected with each other. The risk that the acquaintances the individuals will try to recruit people who have already been recruited by one of his acquaintances will therefore increase over time and increase the risk that the process will die out before global saturation is reached.

WebRDS may have several advantages over standard offline RDS and other Internet-based sampling methods for hidden groups. In comparison to standard RDS it may allow for representative sampling of hidden groups without geographical limits and can potentially generate larger samples than standard RDS. This would also enable valuable data on variables for which design effects are high [14]. Individuals who for various reasons prefer not to access an RDS survey office physically may additionally be willing to take part in an anonymous web survey. The web-based RDS also entails much lower costs than a standard RDS study. Online networks may also cross social boundaries more often than offline social networks (decreased homophily) and are likely to generally produce larger average personal network sizes than offline networks, both of which can decrease design effects [15]
Current online recruitment of hidden groups is based on self-selected samples of persons who access certain Internet sites and click banner ads for a study. These surveys often have participation rates of a few in a thousand to a few in a hundred [20, 21]. As compared with such online sampling, successful webRDS is likely to achieve considerably reduced self-selection bias, because sustained recruitment is likely to be highly correlated to high participations rates.

In summary, we developed a webRDS system to sample men who have sex with men in Vietnam and showed that is was possible to survey participants on a range of sensitive issues, including sexual behavior, while sustaining recruitment and achieving equilibrium. The results indicate that the method could potentially be implemented at a low cost among Internet-using MSM. With further evaluation and among suitable population groups, Internet-based RDS could become a promising method for representative sampling online.

# Figures and Legends

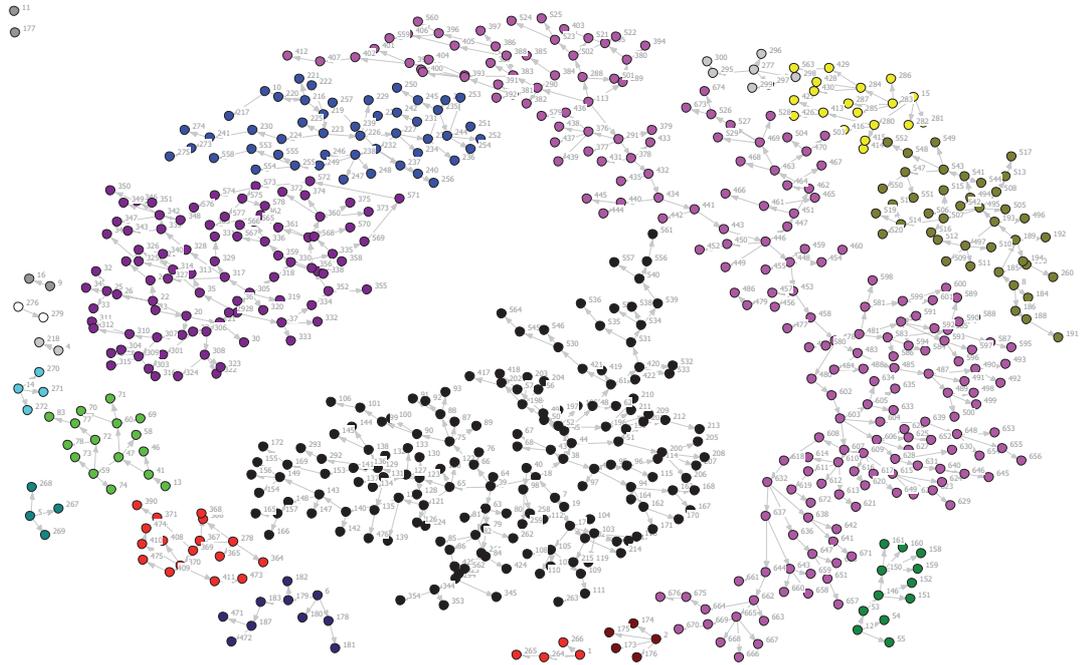

*Figure 1a: The recruitment trees of submitted surveys. Each color represents a separate recruitment tree. Two seeds did not generate further participants.*

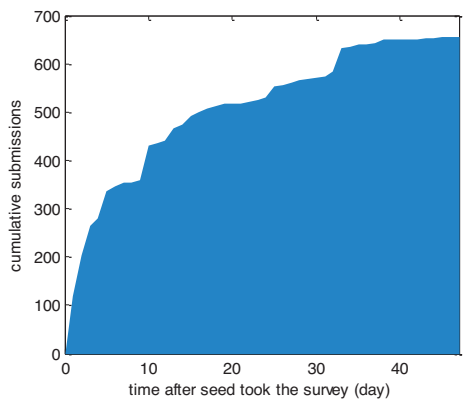

*Figure 1b: Cumulative number of survey submissions over time.*

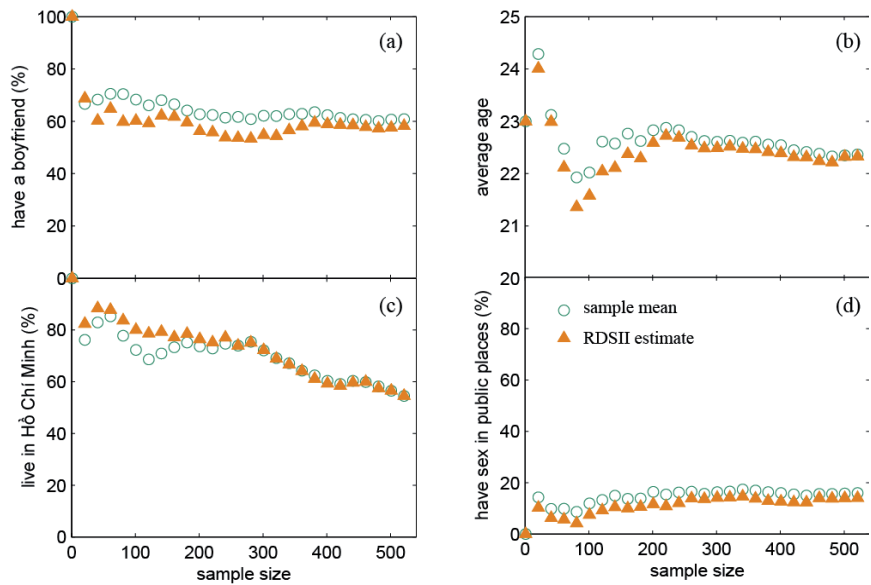

*Figure 2: Change in sample composition with increasing sample size (not adjusted for network size): a) proportion that currently has a romantic relationship; b) average age; c) proportion living in Ho Chi Minh City (the only variable that did not stabilize); and d) proportion who had sex in a public place during the past six months. All variables available in the supplemental material.*

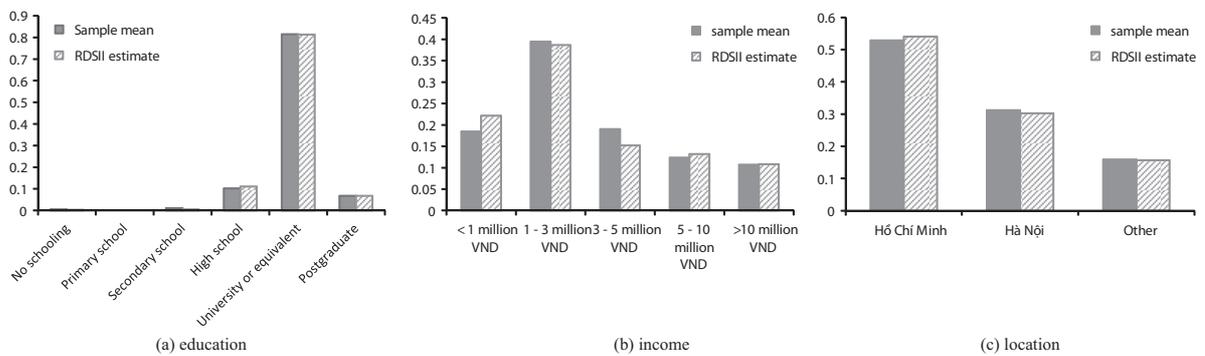

*Figure 3: Sample proportions and estimated population proportions for a) education, b) income, and c) province.*

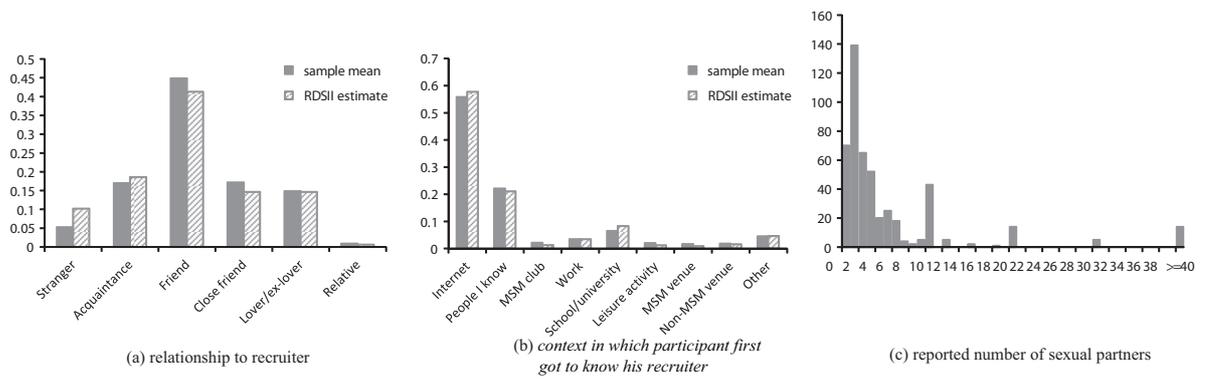

*Figure 4: a) Relationship to recruiter, b) context in which participant first got to know his recruiter, and c) histogram of number of sex partners.*

# Implementation of Web-Based Respondent-Driven Sampling among Men who Have Sex with Men in Vietnam

*Supporting Information*

## 1. RDS estimates for sample data with progressively stricter criteria

We checked all surveys for other signs of duplication or invalid submission. We flagged surveys containing a repeated IP number, deviating answers, or short completion times. We analyzed the sensitivity of the estimates to inclusion and exclusion of these flagged submissions. Specifically, we compared the RDSII estimates generated from the full sample of non-seed submissions with valid age with the RDSII estimates generated from groups with progressively stricter inclusion criteria according to the following:

**Non-Strict:**
All non-seed respondents with valid age (≥18);

**Cleaned sample[1]:**
All non-seed respondents with valid age (≥18);
Exclude submissions with repeated email, Yahoo! Chat ID or telephone number;

**Strict:**
All non-seed respondents with valid age (≥18);
Exclude submissions with repeated email, Yahoo! Chat ID, telephone number;
Exclude submissions with repeated IP address;

**Very Strict :**
All non-seed respondents with valid age (≥18);
Exclude submissions with repeated email, Yahoo! Chat ID, telephone number;
Exclude submissions with repeated IP address;
Exclude submissions with short completion times (< 3 minutes), submissions stating no education (rare in Vietnam), or submission stating six-month partner numbers of above 1,000.

RDSII estimates of all 17 questions for the above sample groups with progressively stricter inclusion criteria are listed in Table 1. The differences between groups are small. The average absolute differences in proportional estimates when comparing the full sample (non-strict criteria) to the other groups, is less than 0.64% (maximum difference 6.6%), and the average absolute differences in numeric estimates is 0.12 (maximum difference 0.30), see Table 1.

---
[1] This is the cleaned sample discussed in the paper.

Table 1: RDSII estimates for samples with progressively stricter inclusion criteria[*]

| Variable | Non-Strict (n=634) | Cleaned sample (n=571) | Strict (n=490) | Very Strict (446) |
|---|---|---|---|---|
| **Proportional estimates (%)** | | | | |
| 1. Have a boyfriend now | 57.5 | 56.6 | 58.0 | 57.4 |
| 2. Longest relationship with men ≥ 6 months | 52.0 | 52.5 | 51.4 | 51.3 |
| 3. Prefer "good looking" when looking for someone for sex | 47.6 | 48.5 | 49.4 | 48.6 |
| 4. Prefer "faithful" when looking for someone for long-term relationship | 39.2 | 41.1 | 41.5 | 39.5 |
| 5. Support same sex marriage | 77.3 | 79.0 | 80.0 | 81.2 |
| 6. Prefer only men as sexual partners | 68.0 | 68.3 | 67.3 | 66.2 |
| 7. Had sex in public places during past 6 months | 14.2 | 13.4 | 11.8 | 12.2 |
| 8. Have education equal to or higher than vocational training after highschool/university | 87.6 | 88.1 | 87.2 | 86.5 |
| 9. Monthly income ≥ 5 million VND | 24.0 | 24.0 | 25.7 | 25.9 |
| 10. Live in Hồ Chí Minh city | 53.4 | 54.1 | 55.3 | 55.9 |
| 11. Use the Internet everyday | 57.0 | 59.6 | 63.6 | 62.2 |
| 12. Recruited by friend | 40.9 | 41.3 | 42.9 | 42.3 |
| 13. First go to know recruiter through friends, lovers or relatives | 21.4 | 21.1 | 20.0 | 20.3 |
| **Numeric estimates** | | | | |
| 14. Number of men had sex with during past 6 months | 4.03 | 4.07 | 3.96 | 4.05 |
| *15. Average age* | 22.23 | 22.22 | 22.07 | 22.06 |
| *16. Number of MSM friends* | 6.74 | 6.76 | 6.91 | 7.04 |
| *17. Number of MSM friends who use Internet* | 5.23 | 5.24 | 5.43 | 5.47 |

*For categorical questions one answer per question is shown

## 2. Equilibrium curves for all variables surveyed in study

To get an overview of whether the sample reached equilibrium, we plot both the sample proportion and the RDSII estimates along with the increased sample size for all variables surveyed in this study (see Figure 1).

To measure the change during the last part of the sampling process when the sampling compositions should have stabilized, we calculate the changes in the sample compositions comparing the full sample and the full sample excluding the last 200$^{th}$ respondents:

$$\Delta p = \left| \Delta p_n - \Delta p_{n-200} \right| \qquad (1.1)$$

$$\Delta \hat{p}^{RDSII} = \Delta \hat{p}_n^{RDSII} - \Delta \hat{p}_{n-200}^{RDSII} \qquad (1.2)$$

Or if the answer is numeric (e.g., age, number of sexual partners, number of friends):

$$\Delta \hat{y}^{RDSII} = \Delta \hat{y}_n^{RDSII} - \Delta \hat{y}_{n-200}^{RDSII} \qquad (1.3)$$

We can see that except for answers for province of residence, the sample proportion and RDS estimates for all other variables became quite stable after 100~200 submissions, the maximum absolute difference between last 200 respondents, is 5.3% for raw sample proportions, 4.3% for RDSII estimated proportions, and 0.67 for the RDSII estimated average numerical variables, i.e., $\max\{\Delta p\} = 5.3\%$, $\max\{\Delta \hat{p}^{RDSII}\} = 5.3\%$, $\max\{\Delta \hat{y}^{RDSII}\} = 0.67$.

The decreased proportions of respondents from Hồ Chí Minh City, indicates that the recruitment chains on average spread from Hồ Chí Minh City to other provinces. The stable estimates for variables not related to place of residence suggest that there is little difference between provinces regarding the social and sexual behaviors/opinions studied here.

There may be a number of possible reasons for not reaching stable proportions for place of residence in the present study sample. Internet-using persons from countryside may have less access to computers in the home and thus take longer time before completing surveys. The network may also be less developed between the large cities and the countryside and social networks between any two rural provinces may be less developed than between rural and urban provinces causing recruitments between rural provinces to go through urban hubs.

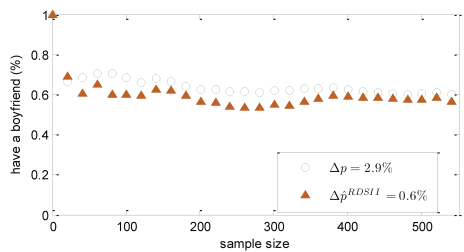
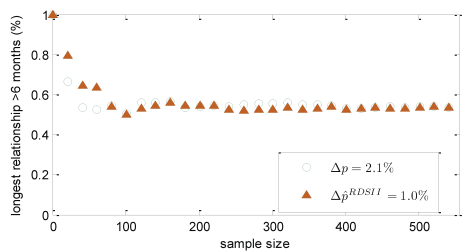
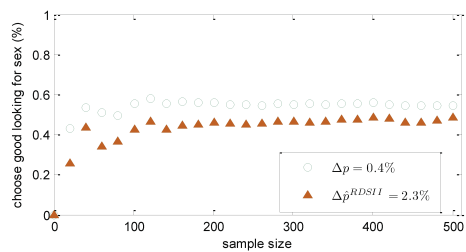
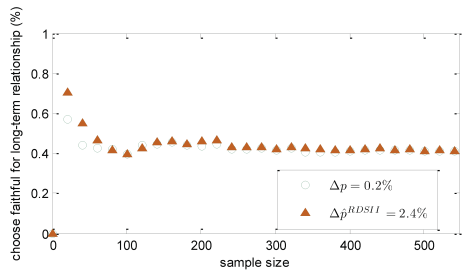
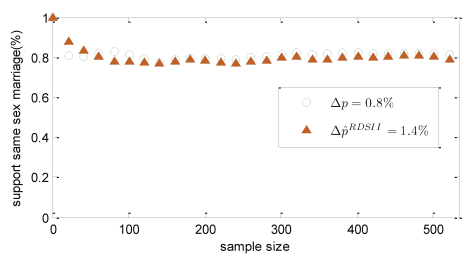
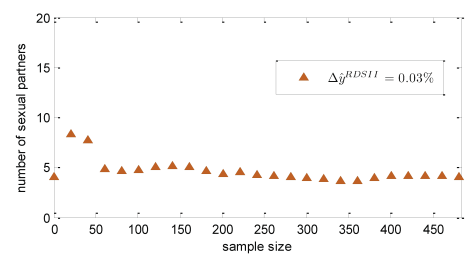
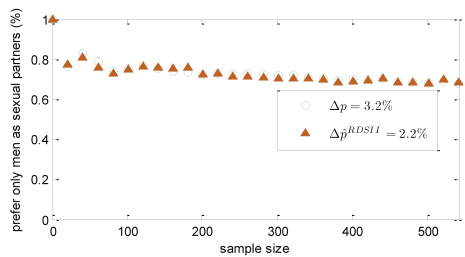
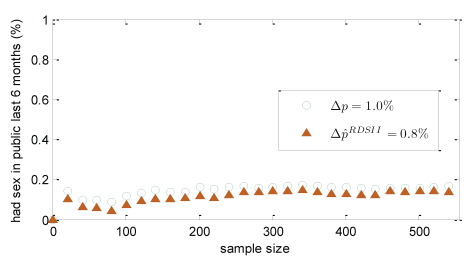
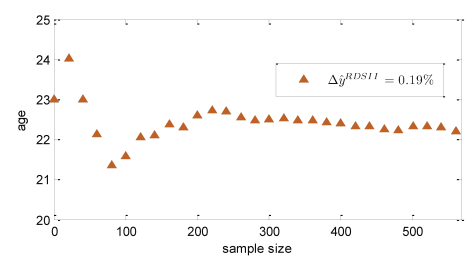

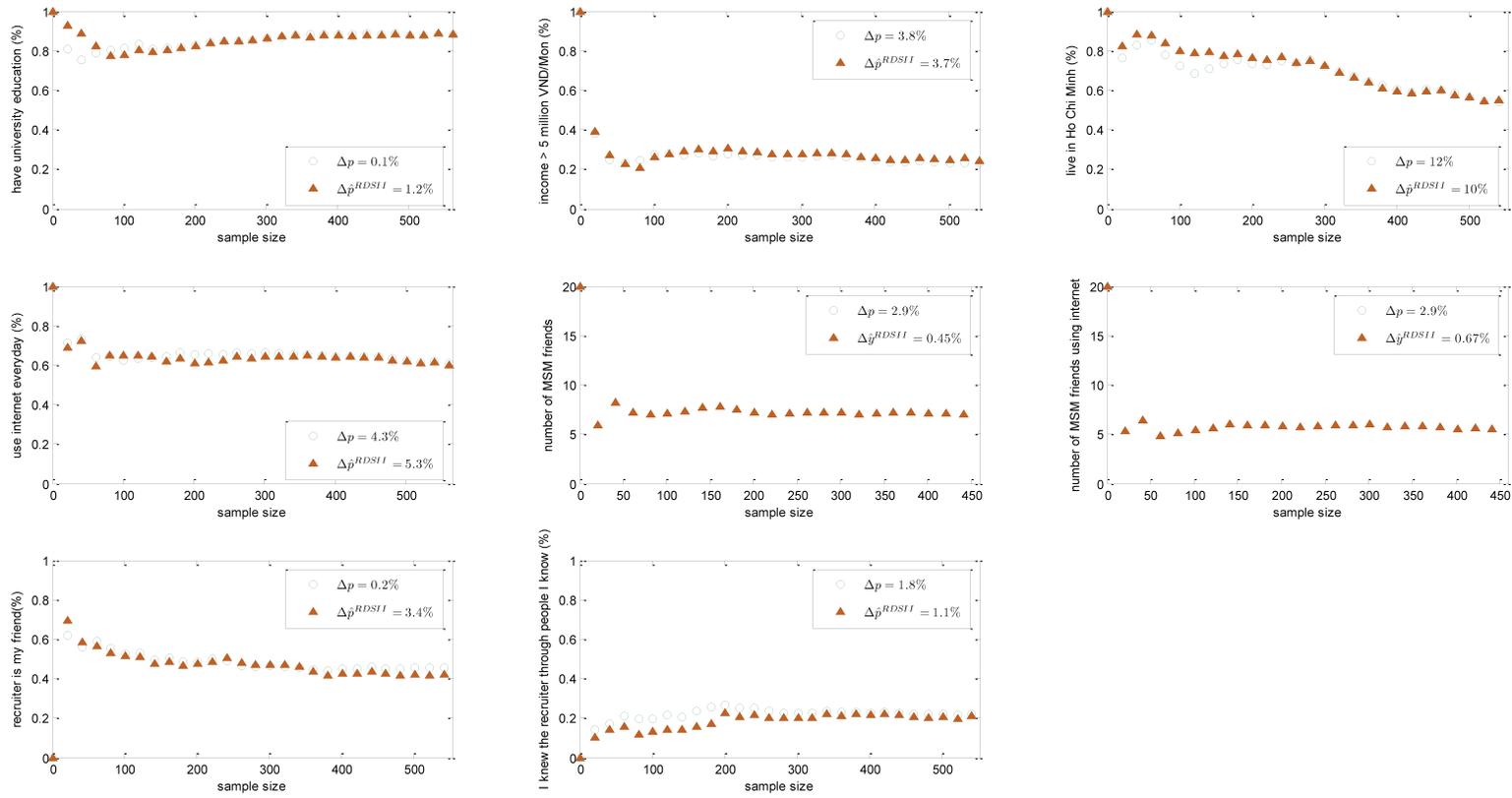

**Figure 1: Sample proportions and RDSII estimates with increased sample size**[2]

---

[2] Cleaned sample used. The curves for sample groups with other criteria are similar.